\documentclass[fleqn,12pt,twoside]{article}
\usepackage{espcrc1}

\usepackage{graphicx}
\usepackage{epsfig}
\usepackage[figuresright]{rotating}

\newcommand{\AmS}{{\protect\the\textfont2
  A\kern-.1667em\lower.5ex\hbox{M}\kern-.125emS}}

\newcommand{\bnl}           {$\rm^{1}$}
\newcommand{\ires}          {$\rm^{2}$}
\newcommand{\kraknuc}       {$\rm^{3}$}
\newcommand{\krakow}        {$\rm^{4}$}
\newcommand{\baltimore}     {$\rm^{5}$}
\newcommand{\newyork}       {$\rm^{6}$}
\newcommand{\nbi}           {$\rm^{7}$}
\newcommand{\texas}         {$\rm^{8}$}
\newcommand{\bergen}        {$\rm^{9}$}
\newcommand{\bucharest}     {$\rm^{10}$}
\newcommand{\kansas}        {$\rm^{11}$}
\newcommand{\oslo}          {$\rm^{12}$}

\title{\bf The New Physics at RHIC.\\ From Transparency to High p$_t$
Suppression.}

\author{
  J. J. Gaardh{\o}je\nbi ~for the BRAHMS Collaboration\\
  \vspace{1mm}
 I.~Arsene\bucharest,
  I.~G.~Bearden\nbi, 
  D.~Beavis\bnl, 
  C.~Besliu\bucharest, 
  B.~Budick\newyork, 
  H.~B{\o}ggild\nbi, 
  C.~Chasman\bnl, 
  C.~H.~Christensen\nbi, 
  P.~Christiansen\nbi, 
  J.~Cibor\kraknuc, 
  R.~Debbe\bnl, 
  E. Enger\oslo,  
  J.~J.~Gaardh{\o}je\nbi, 
  M.~Germinario\nbi, 
  K.~Hagel\texas, 
  O.~Hansen\nbi, 
  A.~Holm\nbi, 
  H.~Ito\bnl$^,$\kansas, 
  A.~Jipa\bucharest, 
  F.~Jundt\ires, 
  J.~I.~J{\o}rdre\bergen, 
  C.~E.~J{\o}rgensen\nbi, 
  R.~Karabowicz\krakow, 
  E.~J.~Kim\bnl, 
  T.~Kozik\krakow, 
  T.~M.~Larsen\oslo, 
  J.~H.~Lee\bnl, 
  Y.~K.~Lee\baltimore, 
  S.~Lindal\oslo, 
  R.~Lystad\bergen, 
  G.~L{\o}vh{\o}iden\oslo, 
  Z.~Majka\krakow, 
  A.~Makeev\texas, 
  B.~McBreen\bnl, 
  M.~Mikelsen\oslo, 
  M.~Murray\texas$^,$\kansas, 
  J.~Natowitz\texas, 
  B.~Neumann\kansas, 
  B.~S.~Nielsen\nbi, 
  J.~Norris\kansas, 
  D.~Ouerdane\nbi, 
  R.~P\l aneta\krakow, 
  F.~Rami\ires, 
  C.~Ristea\bucharest, 
  O.~Ristea\bucharest, 
  D.~R{\"o}hrich\bergen, 
  B.~H.~Samset\oslo, 
  D.~Sandberg\nbi, 
  S.~J.~Sanders\kansas, 
  R.~A.~Scheetz\bnl, 
  P.~Staszel\nbi$^,$\krakow, 
  T.~S.~Tveter\oslo, 
  F.~Videb{\ae}k\bnl, 
  R.~Wada\texas, 
  Z.~Yin\bergen, 
  I.~S.~Zgura\bucharest\\ [1ex]
  \bnl~Brookhaven National Laboratory, Upton, New York 11973, USA\\
  \ires~Institut de Recherches Subatomiques and Universit{\'e} Louis
  Pasteur, Strasbourg, France\\ \kraknuc~Institute of Nuclear Physics,
  Krakow, Poland\\ \krakow~M. Smoluchkowski Inst. of Physics,
  Jagiellonian University, Krakow, Poland\\ \baltimore~Johns Hopkins
  University, Baltimore 21218, USA\\ \newyork~New York University, New
  York 10003, USA\\ \nbi~Niels Bohr Institute,
  University of Copenhagen, Copenhagen 2100, Denmark\\ \texas~Texas
  A$\&$M University, College Station, Texas, 17843, USA\\
  \bergen~University of Bergen, Department of Physics, Bergen,
  Norway\\ \bucharest~University of Bucharest, Romania\\
  \kansas~University of Kansas, Lawrence, Kansas 66045, USA \\
  \oslo~University of Oslo, Department of Physics, Oslo, Norway\\
 }

\begin{document}

\maketitle

\begin{abstract}
  Heavy ion collisions at RHIC energies (Au+Au
  collisions at $\sqrt{s_{NN}}=200$ GeV) exhibit significant new features as compared
  to earlier experiments at lower energies. The reaction is characterized by a high degree of
  transparency of the collisions partners leading to the formation of a baryon-poor central
  region. In this zone, particle production occurs mainly from the stretching of the color
  field. The initial energy density is well above the one considered necessary
  for the formation of the Quark Gluon Plasma, QGP. The production of charged particles
  of various masses is consistent with chemical and thermal equilibrium. Recently, a suppression
  of the high transverse momentum component of hadron spectra has been observed in central
  Au+Au collisions. This can be explained by the energy loss experienced by leading partons
  in a medium with a high density of unscreened color charges. In contrast, such high $p_t$ jets
  are not suppressed in d+Au collisions suggesting that the high $p_t$ suppression is not
  due to initial state effects in the ultrarelativistic colliding
  nuclei.

\end{abstract}

\section{Introduction}
With the Relativistic Heavy Ion Collider, RHIC, at Brookhaven
National Laboratory a new chapter in the story of the production
and study of nuclear matter at extreme energy density and
temperature is being written. In collisions between gold nuclei at
100AGeV+100AGeV at RHIC the total energy in the center of mass is
almost 40TeV, the largest so far achieved in nucleus-nucleus
collisions under laboratory conditions. This energy is so large
that if a sizeable fraction of the initial kinetic energy can be
converted into matter production, many thousands of particles can
be created in a limited volume. It is expected that a region of
matter consisting of free quarks and gluons may be formed in the
early stages of the collision, the so called Quark Gluon Plasma
(QGP). The salient questions to answer are: how much of the
initial (longitudinal) kinetic energy is converted to other
degrees of freedom, what is the energy density achieved and are
there specific signals that carry evidence for the nature of the
formed high density state (i.e. is it partonic or hadronic)? In
this talk I will try to answer some of these questions based on
the results from the first rounds of experiments with RHIC.

RHIC started beam operations in the summer of year 2000 with a
short commissioning run colliding Au nuclei at $\sqrt{s_{NN}}=130$
GeV). The first full run at top energy ($\sqrt{s_{NN}}=200$ GeV)
took place in the fall/winter of 2001/2002. The third RHIC run
during the winter/spring of 2003 focussed on d+Au and p+p
reactions. Here, I concentrate on results from the BRAHMS
detector, one of the four major detectors at RHIC, from the two
long production runs. BRAHMS is a two arm magnetic spectrometer
with excellent momentum resolution and particle identification
capabilities for hadrons. The spectrometers subtend only a small
solid angle (a few msr) but they can rotate in the horizontal
plane about the collision point enabling the collection of data on
hadron production over a wide rapidity range (0-4), a unique
feature among the RHIC experiments. Details about the BRAHMS
detector system may be found in~\cite{BRAHMSNIM,BRAHMSmult}.

\section{Transparency at RHIC}

The stopping of colliding nuclei is a long standing subject. In
reactions at energies in the MeV domain atomic nuclei behave like
droplets. In gentle collisions, nuclei can fuse forming an excited
compound system at rest in the CM frame. In somewhat more violent
collisions the droplets may splash and fragment with approximately
isotropic distributions in the CM. Even at energies of several GeV
pr. nucleon, f.for example in reactions with Au ions at the AGS,
where the energy in the nucleon nucleon center of mass system is
$\sqrt{s_{NN}}=5$ GeV, the colliding nuclei, as seen from the CM
frame, experience significant stopping and a large fraction of the
initial kinetic energy is converted into excitations which
subsequently decay and release energy for example in a direction
transverse to the beam direction.

\begin{figure}[ht]
\begin{center}
  \epsfig{file=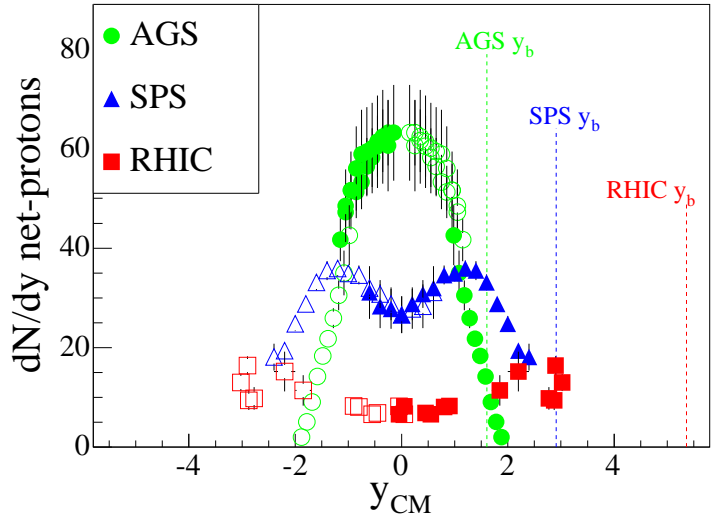,width=10cm}
  \vspace{-7mm}
  \caption{Preliminary rapidity densities of net protons (i.e. number of protons
  minus antiprotons)
  measured at AGS, SPS, and RHIC(BRAHMS). At RHIC, the full distribution
  cannot be measured with current experiments, but BRAHMS will be
  able to extend its measurements to y=3.5 in coming runs,
  corresponding to measurements at 2.3 degrees with respect to the
  beam.}
  \vspace{-7mm}
 \label{fig1}

 \end{center}
\end{figure}

A useful way to quantify the stopping is by the rapidity loss
experienced by the baryons in the colliding nuclei. Rapidity is
defined as
\begin{equation}
y={{1}\over{2}} ln({{E+p_z} \over {E-p_z}})={{1}\over{2}}
ln({{1+\beta cos\theta}\over {1-\beta cos\theta}})
\end{equation}
where $E, pz, \beta$ and $\theta$ denote the total energy,
longitudinal momentum, velocity and angle relative to the beam
axis, respectively, of a particle. If incoming beam baryons have
rapidity, $y_b$ relative to the CM (which has $y=0$) and average
rapidity
\begin{equation}
<y> = \int_0^{y_b} y {{dN}\over{dy}} dy
\end{equation}
after the collision, the rapidity loss is $\delta y = y_b - <y>$.
Here dN/dy denotes the number of particles per unit of rapidity.
Thus, for the extreme case of full stopping: $\delta y = y_b$.
This corresponds to the situation found at very low energies where
all the beam baryons loose all their kinetic energy. In the
expression above a complication arises at CM energies large enough
to allow for the formation of baryon-antibaryon pairs. Thus the
baryon dN/dy distribution to be used is that for the net number of
baryons ( i.e. the difference between the number of baryons and
antibaryons).

\begin{figure}[ht]
\begin{center}
 \epsfig{file=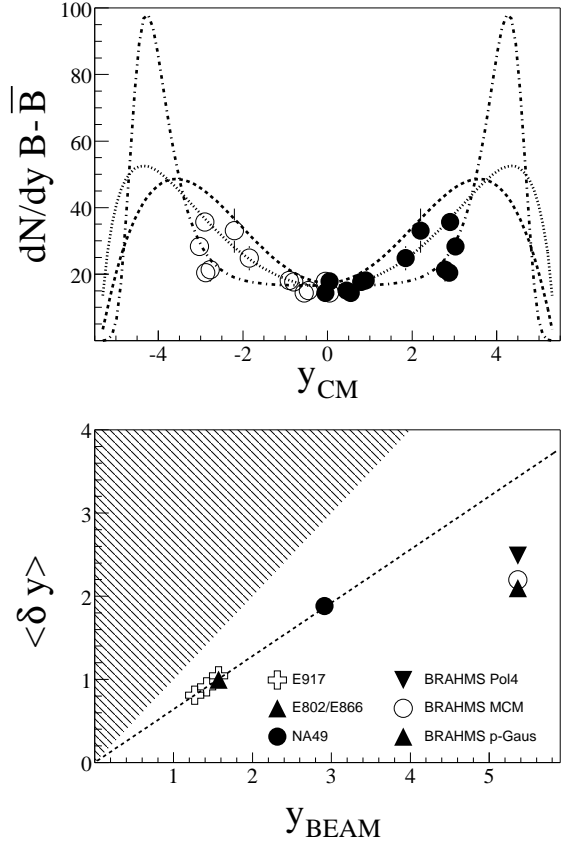,width=8cm}
 \vspace{-9mm}
  \caption{Upper panel: estimates of possible net-baryon distributions requiring
  baryon number conservation. We have assumed that $N(n)\approx N(p)$ and
  scaled hyperon yields at midrapidity to forward rapidity using HIJING.
  From these extremes, limits on the rapidity
  loss of colliding Au ions at RHIC can be set (lower panel). The BRAHMS data are preliminary.}
 \label{fig2}
 \vspace{-9mm}
 \end{center}
\end{figure}

At AGS the number of produced antibaryons is quite small and the
the net-baryon distribution is similar to the proton distribution.
The net-proton rapidity distribution is centered around $y=0$ and
is rather narrow. The rapidity loss is about 1 for a beam rapidity
of approx. 1.6. At CERN-SPS energies ($\sqrt{s_{NN}}=17$ GeV, 158
AGeV Pb+ Pb reactions) the rapidity loss is about 2 for a beam
rapidity of 2.9, about the same relative rapidity loss as at AGS.
The fact that the rapidity loss is large on an absolute scale
means, however, that there is still a sizeable energy loss of the
colliding nuclei. This energy is available for particle production
and other excitations. Indeed, in collisions at SPS,
multiplicities of charged hadrons are about dN/dy=180 around y=0.
At SPS another feature is visible (see fig. 1): the net proton
rapidity distribution shows a double 'hump' with a dip around y=0.
This is a consequence of two effects: the finite rapidity loss of
the colliding nuclei and the finite width of each of the humps,
which reflect the rapidity distributions of the protons in the
colliding nuclei after the collisions. This picture suggests that
the reaction at SPS is beginning to be transparent in the sense
that fewer of the original baryons are found at midrapidity after
the collisions, in contrast to the situation at lower energies.

We have measured the net proton rapidity distribution at RHIC in
the interval $y=0-3$ in the first run with (0-10\%) central Au+Au
collisions at full energy. The beam rapidity at RHIC is about 5.4.
Details of the analysis may be found in
refs.\cite{BRAHMSnetproton} and in the recent Ph.D. thesis of P.
Christiansen. The results are displayed in fig.~\ref{fig1}
together with the previously discussed net-proton distributions
measured at AGS and SPS. It is notable that the RHIC distribution
is both qualitatively and quantitatively different from those at
lower energies. The net number of protons per unit of rapidity
around y=0 is only about 7 and the distribution is flat over at
least the first unit of rapidity. The distribution increases in
the rapidity range $y=2-3$ to an average $dN/dy \approx 12$. We
have not yet completed the measurements at the most forward angles
(highest rapidity) allowed by the geometrical setup of the
experiment, but we can exploit that there must be baryon
conservation in the reactions to try to set limits on the relative
rapidity loss at RHIC. This is illustrated in fig.~\ref{fig2},
which shows various possible distributions whose integral areas
correspond to the number of baryons present in the overlap between
the colliding nuclei. From such distributions one may deduce a set
of upper and lower limits for the rapidity loss at RHIC. In
practice the situation is complicated by the fact that not all
baryons are measured. We measure in BRAHMS the direct protons, but
only some of the decay protons from for example $\Lambda$. The
limits shown in the figure include some reasonable estimates of
these effects ~\cite{BRAHMSnetproton,STARPHENIXlambda130}.

The conclusion is that the {\it absolute} rapidity loss at RHIC
$(\delta y =2.2 \pm 0.4)$ is not appreciably larger than at SPS.
In fact the {\it relative} rapidity loss is significantly reduced
as compared to an extrapolation of the low energy
systematics~\cite{FVandOHstopping}. It should be noted that the
rapidity loss is still significant and that, since the overall
beam energy (rapidity) is larger at RHIC than at SPS, the absolute
energy loss increases appreciably from SPS to RHIC thus making
available a significantly increased amount of energy for particle
creation in RHIC reactions.

\begin{figure}[ht]
\begin{center}
  \epsfig{file=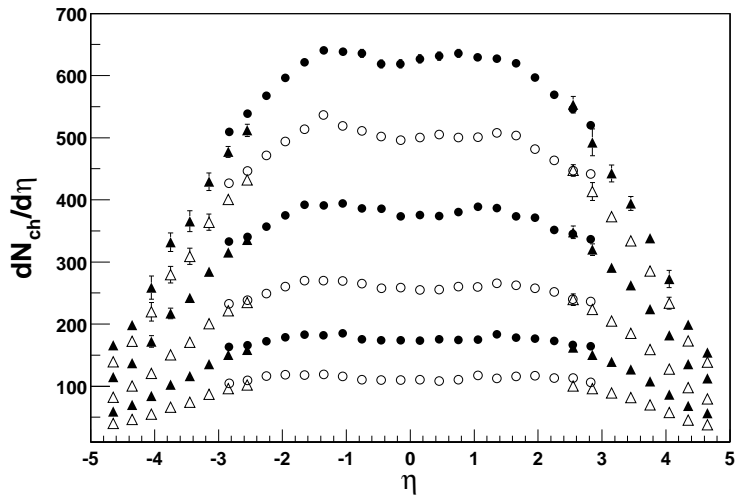,width=10cm}
  \vspace{-7mm}
  \caption{Pseudorapidity densities (multiplicities) of charged particles measured by BRAHMS for
   $\sqrt{s_{NN}}=200$ GeV) Au+Au collsions. The various distributions
   correspond to collisions centralities 0-5\% (top),
   5-10\%,10-20\%,20-30\%,30-40\%, 40-50\%. The integral of the most central distribution
   corresponds to about 4600 charged particles~\cite{BRAHMSmult}.}
 \label{fig3}
 \vspace{-9mm}
 \end{center}
\end{figure}

\section{Particle production and energy density}

The stopping scenario that we observe at RHIC and which was
outlined in the previous section entails that the reaction can be
viewed as quite transparent (opaque is perhaps a better word).
After the collision, the matter and energy distribution can be
conceptually divided up into two main parts, namely a so-called
fragmentation region consisting of the excited remnants of the
colliding nuclei which have experienced an average rapidity loss
of about 2.2 and a central region in which few of the original
baryons are present but where significant energy density is
collected. This picture is consistent with the schematic one
already proposed by Bjorken 20 years ago~\cite{Bjorken83}. The
central region (an interval around midrapidity) is decoupled from
the fragments. The energy removed from the kinetic energy of the
fragments is initially stored in a color field strung between the
receeding partons that have interacted. The linear increase of the
color potential with distance eventually leads to the production
of quark-antiquark pairs. Such pairs may be produced anywhere
between the interacting partons leading to an approximately
uniform particle production as a function of rapidity. In this
picture, the properties of the particle production is also uniform
as a function of rapidity (boost invariance). If the density of
produced quark-antiquark pairs is sufficiently high, the average
distance between them will be low and the binding potential
between the colored objects will be small. The objects will become
asymptotically free and exist in a plasma like state until the
subsequent expansion and lowered density leads to confinement and
hadronization.

\begin{figure}[ht]
\begin{center}
 \epsfig{file=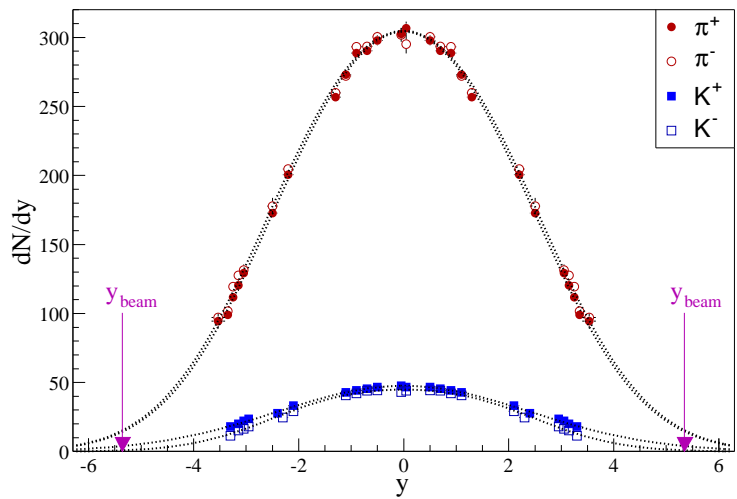,width=10cm}
 \vspace{-7mm}
  \caption{Rapidity density distribution for positive
  and negative pions and kaons. Data points collected at positive y have been
  reflected around y=0. Preliminary BRAHMS data.}
 \label{fig4}
\vspace{-7mm}
\end{center}
\end{figure}

Figure~\ref{fig3} shows the overall multiplicity of charged
particles observed in Au+Au collisions at RHIC
~\cite{BRAHMSmult,PHOBOSmult200} for various collision
centralities and as a function of pseudorapidity (pseudorapidity,
$\eta$, is defined as $\eta=ln(cot(\theta/2))$ and is a customary
rapidity variable for non identified particles). The figure shows
that the multiplicity at RHIC is about $dN/d\eta = 625$ charged
particles pr. unit rapidity around $\eta=0$ for central
collisions. The production of charged particles in central
collisions exceeds the particle production seen in p+p collisions
at the same energy by about 40\%, when the yield seen in p+p
collisions is multiplied by the number of participant nucleon
pairs in the overlap region between the colliding nuclei.

\begin{figure}[ht]
\begin{center}
  \epsfig{file=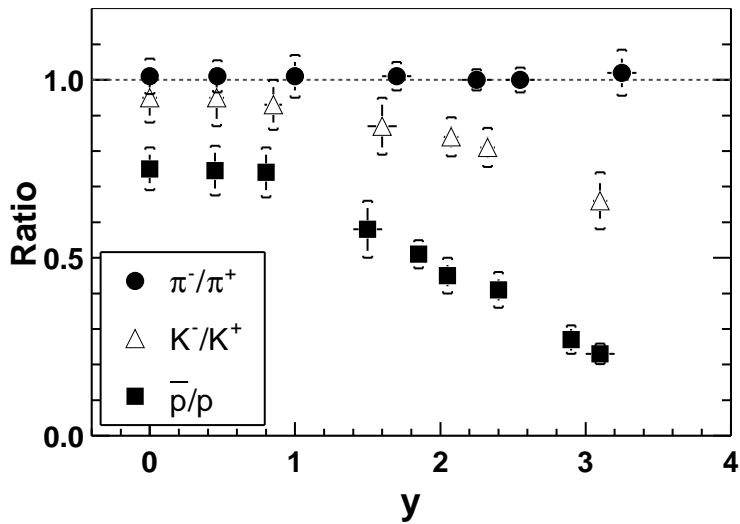,width=10cm}
  \vspace{-9mm}
  \caption{Ratios of antiparticles to particles (pions, kaons and protons)
  as a function of rapidity for $\sqrt{s_{NN}}=200$ GeV Au+Au collisions
  measured by the BRAHMS experiment~\cite{BRAHMSratios}.}
 \label{fig5}
 \vspace{-9mm}
 \end{center}
\end{figure}

\begin{figure}[ht]
\begin{center}
  \epsfig{file=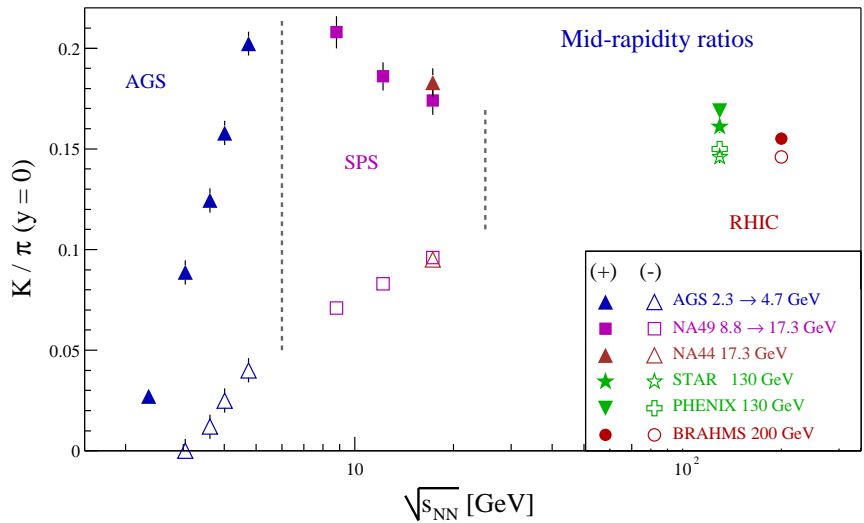,width=12cm}
  \vspace{-7mm}
  \caption{Ratios of kaons and pions of both charge signs as a function of center
  of mass energy in the nucleon-nucleon system at midrapidity. At top RHIC energy
  the two ratios are about the same and equal to 0.15. Preliminary BRAHMS data. }
 \label{jjKoverpi}
 \vspace{-9mm}
 \end{center}
\end{figure}

Figure~\ref{fig4} shows a recent and more detailed study of the
particle production in central collisions as a function of
rapidity (the PhD. work of Djamel Ouerdane~\cite{DjamelPhD}). The
figure shows the rapidity densities of pions and kaons for central
collisions. From such distributions we can construct the ratio of
the yields of particles and their antiparticles as a function of
rapidity. Figure~\ref{fig5} shows the ratios of yields of
antihadrons to hadrons (posititive pions, kaons and protons and
their antiparticles). The ratio is seen to be approaching unity in
an interval of about 1.5 units of rapidity around midrapidity,
suggesting that the particle production is predominantly from pair
creation. This is exactly true for pions (ratio of 1), but less so
for kaons (ratio= 0.95) and protons (ratio= 0.76). The reason is
that there are other processes that break the symmetry between
particles and antiparticles that depend on the net-baryon
distribution discussed in the previous section. One such process
that is relevant for kaons is the associated production mechanism
(e.g. $p+p \rightarrow p + \Lambda + K^+$) which leads to an
enrichment of positive kaons in regions where there is an excess
of baryons. Support for this view is given by
fig.~\ref{jjKoverpi}, which shows the systematics of kaon
production relative to pion production as a function of center of
mass energy. At AGS, where the net proton density is high at
midrapidity, the rapidity density of $K^+$ strongly exceeds that
of $K^-$. In contrast, at RHIC, production of $K^+$ and $K^-$ is
almost equal. This situation changes, however, at larger
rapidities where the net proton density increases.

Integration of the charged particle pseudorapidity distributions
corresponding to central collisions tells us that about 4600
charged particles are produced in each of the 5\% most central
collisions. Since we only measure charged particles, which are
predominantly pions and kaons, as may be seen from
fig.~\ref{fig4}, and not the neutrals, we multiply this
multiplicity by 3/2 to obtain the total particle multiplicity of
about 7000 particles.

\begin{figure}[ht]
\begin{center}
 \epsfig{file=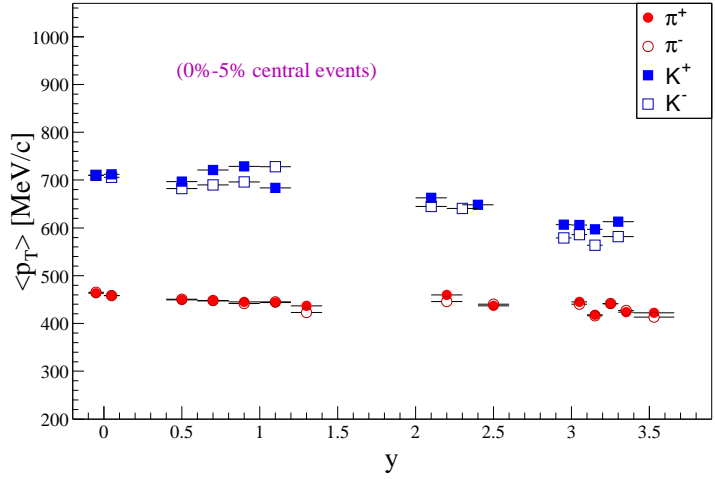,width=10cm}
\vspace{-7mm}
  \caption{Distribution of average transverse momenta of pions and
  kaons as a function of rapidity as measured by BRAHMS. Preliminary.}
 \label{fig6}
 \vspace{-9mm}
\end{center}
\end{figure}

From the measured spectra of pions, kaons and protons and their
antiparticles  as a function of transverse momentum we can
determine the average transverse momentum for each particle
species (fig.~\ref{fig6}). This allows us to estimate the initial
energy density from Bjorkens formula:
\begin{equation}
\epsilon = {{1} \over {\pi R^2 \tau}} {{d<E_t>} \over {d\eta}}
\end{equation}
where we can make the substitution $d<E_t> = <m_t> dN$ and use
quantities from the measured spectral distributions. Since we wish
to calculate the energy density in the very early stages of the
collision process we may use for R the radius of the overlap disk
between the colliding nuclei, thus neglecting transverse
expansion. The formation time is more tricky. It is often assumed
to be of the order of 1 fm/c, but it might be shorter for heavier
or more energetic particles, as the uncertainty relation would
tell us. Under these assumptions we find that $\epsilon > 5
GeV/fm^3$. This value of the initial energy exceeds by a factor of
30 the energy density of a nucleus, by a factor of 10 the energy
density of a baryon and by a factor of 5 the critical energy
density for QGP formation that is predicted by lattice QCD
calculations.

\section{Is there thermodynamical and chemical equilibrium at
RHIC?}

\begin{figure}[ht]
\begin{center}
  \epsfig{file=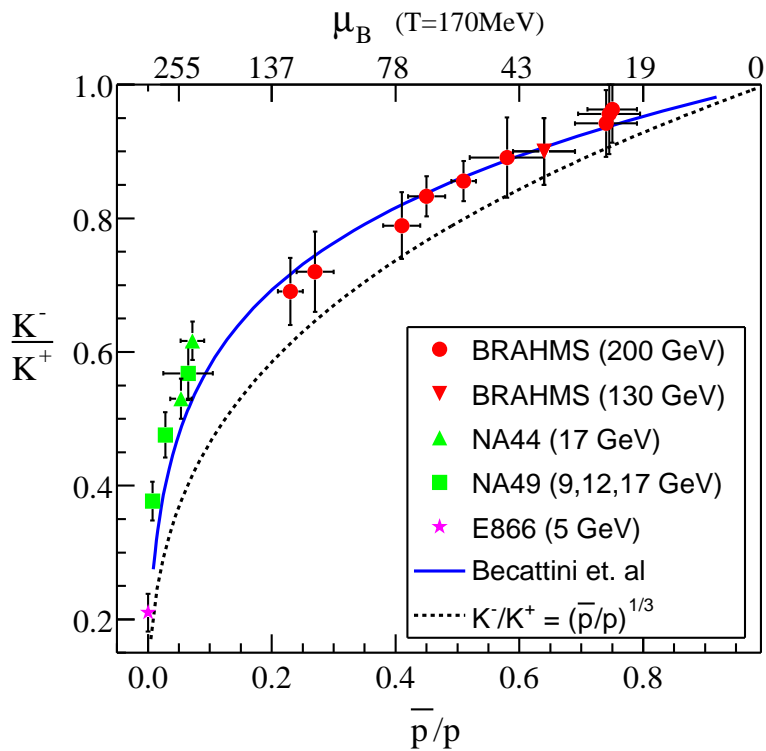,width=10cm}
  \vspace{-7mm}
  \caption{Correlation between the ratio of charged kaons and the ratio of
  antiprotons to protons. The dashed curve corresponds to equation 3 in the text.
  The full drawn curve is a statistical model calculation with a
  chemical freeze-out temperature of 177MeV.}
 \label{fig7}
 \vspace{-9mm}
 \end{center}
\end{figure}

The particle yields measured by BRAHMS also lend themselves to an
analysis of the charged particle production in terms of the
statistical
model~\cite{BRAHMSratios},\cite{Koch86,Cleymans93,Cleymans94,PBM01statmodel}.
Figure~\ref{fig7} shows the ratios of negative kaons to positive
kaons as a function of the corresponding ratios of antiprotons to
protons for various rapidities at RHIC. The data are for central
collisions, and the figure also displays similar ratios for heavy
ion collisions at AGS and SPS energies. There is a striking
correlation between the RHIC/BRAHMS kaon and proton ratios over 3
units of rapidity. Assuming that we can use statistical arguments
based on chemical and thermal equilibrium at the quark level, the
ratios can be written
\begin{equation}
{{\rho(\bar{p})}\over{\rho(p)}}= exp({{-6\mu_{u,d}}\over{T}})
\end{equation}
and
\begin{equation}
{{\rho(K^-)}\over{\rho(K^+)}}= exp({{-2(\mu_{u,d}
-\mu_{s})}\over{T}})
          = exp({{2\mu_s}\over{T}})\times
          [{{\rho(\bar{p})}\over{\rho(p)}}]^{{1}\over{3}}
\end{equation}
where $\rho, \mu$ and $T$ denote number density, chemical
potential and temperature, respectively. From equation 4 we find
the chemical potential for u and d quarks (often called the
baryochemical potential) to be around 25 MeV, the lowest value yet
seen in nucleus-nucleus collisions. Equation 5 tells us that for a
vanishing strange quark chemical potential we would expect a power
law relation between the two ratios with exponent 1/3. The
observed correlation is well described by the relationship
$\rho(K^-)/\rho(K^+)= \rho(\bar p)/\rho(p)^{0.24}$, i.e. with an
exponent that is close to 1/4 suggesting, a finite value of the
strange quark chemical potential. A more elaborate analysis for a
grand canonical ensemble assuming charge, baryon and strangeness
conservation can be carried out by fitting these and many other
particle ratios observed at RHIC in order to obtain the chemical
potentials and temperatures. An example of such a procedure is
shown in fig.~\ref{fig7} and displayed with the full
line~\cite{Becattini}. Here the temperature is 170MeV. The point
to be made here is that the calculation agrees with the data over
a wide energy range (from SPS to RHIC) and over a wide range of
rapidity at RHIC. This may be an indication that the system is in
chemical equilibrium over the considered $\sqrt{s}$ and $y$ ranges
(or at least locally in the various y bins). Separate measurements
at RHIC of, for example, elliptical flow also point to local
equilibration around midrapidity.

\section{High pt suppression. The smoking gun of QGP?}

The discussion in the previous sections indicates that the
conditions for particle production in a interval $|y| < 1.5-2$ are
radically different than in reactions at lower energies. At RHIC
the central zone is nearly baryon free, the considered rapidity
interval appears to approximately exhibit the anticipated boost
invariant properties, the particle production is large and
dominated by pair production and the energy density appears to
exceed significantly the one required for QGP formation. The
overall scenario is therefore consistent with particle production
from the color field, formation of a QGP and subsequent
hadronization. But is this interpretation unique and can more
mundane explanations based on a purely hadronic scenario be
excluded? In spite of the difficulties in reconciling the high
initial energy density with hadronic volumes, a comprehensive
answer to this question requires the observation of an effect that
is directly dependent on the partonic or hadronic nature of the
formed high density zone.

Such an effect has recently been discovered at RHIC and is related
to the suppression of the high transverse momentum component of
hadron spectra in central Au+Au collisions as compared to scaled
momentum spectra from p+p
collisions~\cite{BRAHMShighpt,STARptAuAu,PHENIXptAuAu,PHOBOSptAuAu}.
The effect, originally proposed by Bjorken, Gyulassy and
others~\cite{Bjorken83,Gyulassy90,WANG92} is based on the
expectation of a large energy loss of high momentum partons
scattered in the initial stages of the collisions in a medium with
a high density of free color charges. According to QCD colored
objects may loose energy by radiating gluons by bremsstrahlung.
Due to the color charge of the gluons, the energy loss is
proportional to the square of the depth of color medium traversed.
Such a mechanism would strongly degrade the energy of leading
partons resulting in a reduced transverse momentum of leading
particles in the jets that emerge after fragmentation into
hadrons. The STAR experiment has shown that the topology of high
pt hadron emission is consistent with jet emission, so that we may
really speak about jet-suppression.

\begin{figure}[ht]
\begin{center}
  \epsfig{file=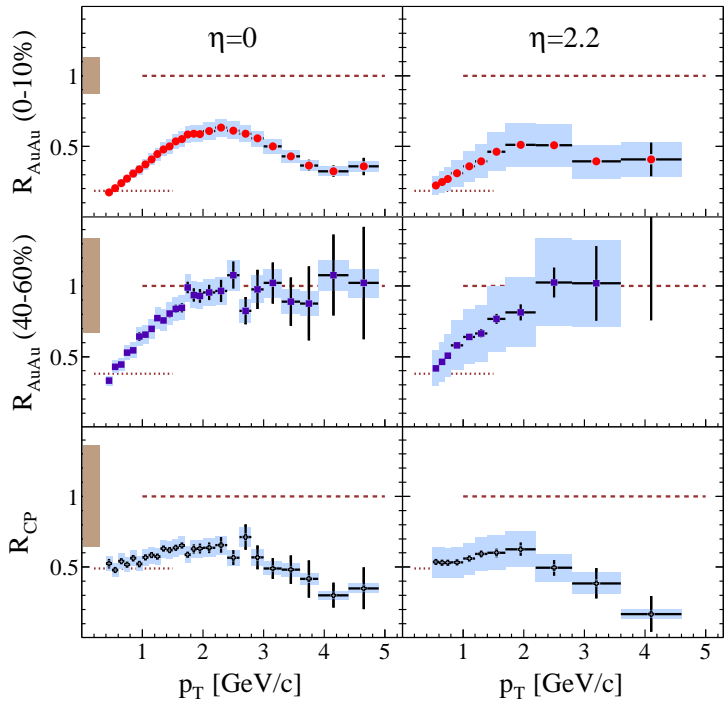,width=10cm}
  \vspace{-7mm}
  \caption{Nuclear modification factors $R_{AuAu}$ as defined in the text,
  for central and semi-peripheral Au+Au collisions at midrapidity (left) and
  forward rapidity (right). The lower row shows the factor $R_{cp}$, i.e the
  ratio of the $R_{AuAu}$ for central and peripheral collisions, which
  has the property of being independent of the p+p reference spectrum.}
\label{fig8}
\vspace{-9mm}
\end{center}
\end{figure}

The two upper rows of fig.~\ref{fig8} show our
~\cite{BRAHMS-qm2002CEJ,BRAHMShighpt} measurements of the
so-called nuclear modification factors for {\em unidentified}
charged hadrons from Au+Au collisions at rapidities $\eta = 0$ and
$2.2$ (this is the ph.d. work of Claus E. J{\o}rgensen, see also
his contribution at this conference). The nuclear modification
factor is defined as:

\begin{equation}
R_{AA}= {{d^2N^{AA}/dp_td\eta}\over{N_{bin}d^2N^{NN}/dp_td\eta}}
\end{equation}

It involves a scaling of measured nucleon-nucleon transverse
momentum distributions by the number of expected incoherent binary
collisions, $N_{bin}$ (see ~\cite{UA1,STARpp}. In the absence of
any modification resulting from the 'embedding' of elementary
collisions in a nuclear collision we expect $R_{AA}=1$ a high
$p_t$. At low $p_t$, where the particle production follows a
scaling with the number of participants, the above definition of
$R_{AA}$ leads to $R_{AA}<1$ for $p_t < 2 GeV/c$. In fact, it is
found that $R_{AA}>1$ for $p_t>2 GeV/c$ in nuclear reactions at
lower energy. This effect, called the Cronin effect, is associated
with initial multiple scattering of high momentum partons.

Figure~\ref{fig8} demonstrates that, surprisingly, $R_{AA}<1$ also
at high $p_t$ for central collisions at both pseudorapidities,
while $R_{AA} \approx 1$ for more peripheral collisions. It is
remarkable that the suppression that is observed at $p_t \approx 4
GeV/c$ is very large, amounting to a factor of 3 for central Au+Au
collisions as compared to $p+p$ and a factor of more than 4 as
compared to the more peripheral collisions. Such large suppression
factors are observed at both pseudorapidities.

\begin{figure}[ht]
\begin{center}
  \epsfig{file=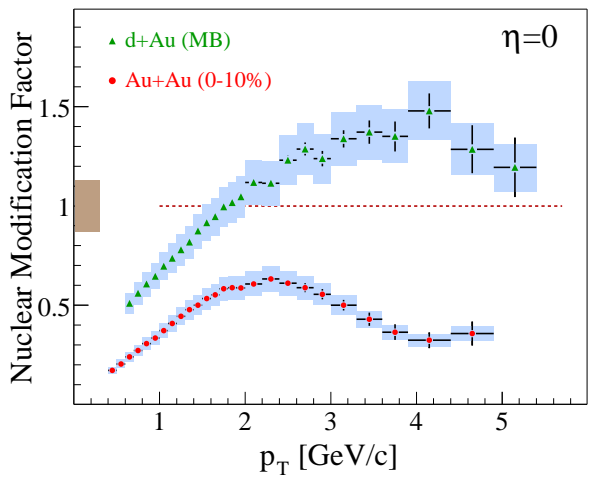,width=8cm}
  \vspace{-7mm}
  \caption{Nuclear modification factors measured for central Au+Au collisions
  and minimum bias d+Au collisions at $\sqrt{s_{NN}}=200$ GeV, evidencing the
  important high pt suppression observed in central Au+Au collisions.}
 \label{fig9}
 \vspace{-10mm}
\end{center}
\end{figure}

It has been conjectured that the observed high $p_t$ suppression
might be the result of an entrance channel effect, for example due
to a limitation in the phase space available for parton collisions
related to saturation effects~\cite{gluonsat} in the gluon
distributions inside the swiftly moving colliding nucleons (which
have $\gamma = 100$). As a test of these ideas we have very
recently determined the nuclear modification factor for 100 AGeV d
+ 100 AGeV Au minimum bias collisions. The resulting $R_{dAu}$ is
shown in fig.~\ref{fig9} where it is also compared to the
$R_{AuAu}$ for central collisions previously shown in
fig.~\ref{fig8}. No high-$p_t$ jet suppression is observed for
d+Au. In fact, the $R_{dAu}$ distribution measured for d+Au shows
the Cronin type enhancement~\cite{Cronin75} observed at lower
energies~\cite{WA98-RAA,NA49-RAA,CERES-RAA}. At $p_t \approx 4
GeV/c$ we find a ratio $R_{dAu}/R_{AuAu} \approx 5$. These
observations are consistent with the smaller transverse dimensions
of the overlap disk between the d and the Au nuclei and also
appear to rule out strong entrance channel effects.

\begin{figure}[ht]
\begin{center}
  \epsfig{file=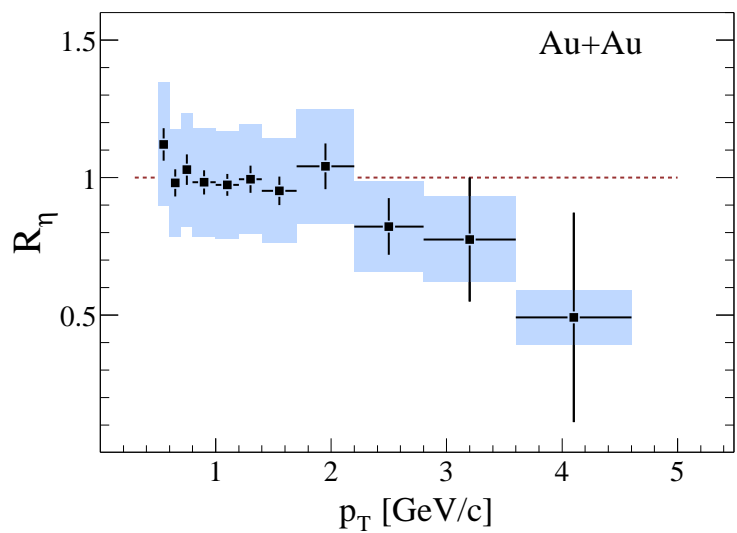,width=10cm}
  \vspace{-7mm}
  \caption{Ratio, $R_\eta$, of the suppression factors $R_{cp}$ at
   pseudorapidities $\eta=0$ and $\eta=2.2$ shown in figure 9. The
   figure suggest that high $p_t$ suppression persists
   (and is even more important) at forward rapidity than at $\eta=0$.}
 \label{fig10}
 \vspace{-10mm}
 \end{center}
\end{figure}

The very large suppression observed in central Au+Au collisions
must be quantitatively understood and will require systematic
dynamic modelling. At $\eta =0$ the particles are emitted at 90
degrees relative to the beam direction, while at $\eta = 2.2$ the
angle is only about 12 degrees. In a naive geometrical picture of
an absorbing medium with cylindrical symmetry around the beam
direction, the large suppression seen at forward angles indicates
that the suppressing medium is extended also in the longitudinal
direction. Since the observed high $p_t$ suppression is similar or
even larger at forward rapidity as compared to midrapidity (see
fig.~\ref{fig10}) one might be tempted to infer a longitudinal
extend of the dense medium which is approximately similar to its
transverse dimensions ($R\approx 5fm$), and from this a life time
longer than $5 fm/c$. However, the problem is more complicated,
due to the significant transverse and in particular longitudinal
expansion that occurs as the leading parton propagates through the
medium, effectively reducing the densities of color charges seen.
There is little doubt that systematic studies of the high
$p_t$-jet energy loss as a function of the thickness of the
absorbing medium obtained by varying the angle of observation of
high $p_t$ jets relative to the beam direction will be required in
order to understand the properties of the dense medium. The BRAHMS
experiment is uniquely suited among the RHIC experiments to carry
out such a {\em QGP tomography}!

\section{Conclusions and perspectives}

The first round of RHIC experiments has clearly shown that we have
moved in high energy nucleus-nucleus collisions into a
qualitatively new physics domain characterized by a high degree of
reaction transparency leading to the formation of a near baryon
free central region. There is, nevertheless, appreciable energy
loss of the colliding nuclei, so the conditions for the formation
of a very high energy density zone in an interval of about
$|dN/dy| < 2\pm 0.5$ around midrapidity are present. The
observation of a very large suppression of high momentum particles
(jets) originating from hard scatterings in the very early stages
of the collision is consistent with a significant energy loss of
high momentum partons moving in a dense medium of unscreened color
charges, which may be the quark gluon plasma. The next round of
experiments at RHIC will afford us the opportunity to go deeper in
determining the properties of the new high density state of matter
that is formed, and disentangling its partonic and hadronic
components~\cite{Vitev03,Gallmeister03}. In particular it is
important to determine and understand the flavor dependence of
high $p_t$ suppression by measuring identified high $p_t$ mesons,
and likewise to compare the suppression of baryonic and mesonic
jets. A systematic scan of the jet suppression as a function of
the length of dense matter traversed can be carried out in both
the transverse and longitudinal directions by varying the system
size and the angle of observation (rapidity). Likewise, a central
future program item must be to determine, through a measurement of
an excitation function, the onset of the jet suppression, which is
conspicuously absent at SPS but strongly present at RHIC.

\section{Acknowledgements}

This work was supported by the Danish Natural Science Research
Council, the division of Nuclear Physics of the Office of Science
of the U.S. DOE, the Research Council of Norway, the Polish State
Committee for Scientific Research and the Romanian Ministry of
Research.

\end{document}